\documentclass[12pt]{article}
\usepackage{jheppub}

\pdfoutput=1

\usepackage{amsmath,bbm,array,amsfonts,graphicx,wrapfig,lscape,float,mathtools,multirow,longtable}
\usepackage[dvipsnames]{xcolor}
\usepackage{array}

\newcommand{\be}{\begin{equation}}
\newcommand{\ee}{\end{equation}}
\newcommand{\beq}{\begin{equation}}
\newcommand{\beql}[1]{\begin{equation}\label{#1}}
\newcommand{\eeq}{\end{equation}}
\newcommand{\ba}{\begin{array}}
\newcommand{\ea}{\end{array}}
\newcommand{\bea}{\begin{eqnarray}}
\newcommand{\beal}[1]{\begin{eqnarray}\label{#1}}
\newcommand{\eea}{\end{eqnarray}}
\newcommand{\ben}{\begin{enumerate}}
\newcommand{\een}{\end{enumerate}}
\newcommand{\bean}{\begin{eqnarray*}}
\newcommand{\eean}{\end{eqnarray*}}

\newcommand{\sref}[1]{\S\ref{#1}}

\newcommand{\fref}[1]{Figure \ref{#1}}
\newcommand{\btab}[1]{\begin{tabular}{#1}}
\newcommand{\etab}{\end{tabular}}

\newcommand{\comment}[1]{}

\newcommand{\qed}{\nobreak \ifvmode \relax \else
      \ifdim\lastskip<1.5em \hskip-\lastskip
      \hskip1.5em plus0em minus0.5em \fi \nobreak
      \vrule height0.75em width0.5em depth0.25em\fi}

\definecolor{darkspringgreen}{rgb}{0.09, 0.45, 0.27}
\definecolor{forestgreen}{rgb}{0.13, 0.55, 0.13}

\usepackage{array}
\usepackage{comment}
\usepackage{physics}
\usepackage{subcaption}
\usepackage{tikz,tikz-3dplot}
\usepackage[colorlinks=true]{hyperref}
\newcolumntype{C}[1]{>{\centering\let\newline\\\arraybackslash\hspace{0pt}}m{#1}}

\definecolor{yellow2}{rgb}{0.98, 0.80, 0.20}

\definecolor{mygreen}{RGB}{24,174,42}

\title{On the Origin of Toric Diagrams}

\author[a,b,c]{Sebasti\'an Franco,}

\author[d,e]{Diego Rodr\'iguez-G\'omez}

\affiliation[a]{Physics Department, The City College of the CUNY\\
	160 Convent Avenue, New York, NY 10031, USA}
\affiliation[b]{Physics Program and \textsuperscript{$c$}Initiative for the Theoretical Sciences\\
	The Graduate School and University Center, The City University of New York\\
	365 Fifth Avenue, New York NY 10016, USA}
\affiliation[d]{Department of Physics, Universidad de Oviedo \\  
C/ Federico Garc\'ia Lorca  18, 33007  Oviedo, Spain}
\affiliation[e]{Instituto Universitario de Ciencias y Tecnolog\'ias Espaciales de Asturias (ICTEA) \\
 C/~de la Independencia 13, 33004 Oviedo, Spain.}

\emailAdd{sfranco@ccny.cuny.edu}
\emailAdd{d.rodriguez.gomez@uniovi.es}

\abstract{Five-dimensional superconformal field theories ($5d$ SCFTs) can be encoded by Generalized Toric Polygons (GTPs), where external legs of the dual $(p,q)$ five-brane web correspond to $T$-cones. Hanany-Witten transitions act on these geometries by flipping $T$-cones about their apex, thereby naturally endowing the choice of origin in the polygon with physical significance. It was recently conjectured that a suitably graded Hilbert series equals the Ehrhart series of the dual polytope, which, in turn, is an invariant under such mutations.
In this paper, we introduce a prescription for assigning scaling dimensions to fields in the toric gauge theory associated with the underlying toric diagram and show that the resulting Hilbert series of the coherent component of the moduli space matches the geometric Hilbert series given by the Ehrhart series of the dual polytope once an origin is specified. We validate our construction through several non-trivial examples, including cases with multiple admissible choices of origin leading to distinct GTPs and brane-web realizations. Our results provide evidence that ordinary brane tilings retain non-trivial information about generalized toric polygons and suggest the existence of a deeper combinatorial structure underlying GTPs.}

\begin{document}

\maketitle

\section{Introduction and conclusions}

Interacting Quantum Field Theories (QFTs) in $5d$ are very interesting, in particular due to the fact that they are intrinsically strongly coupled. For the same reason, they are also notoriously hard to construct. String/M-theory inspired methods offer a useful perspective into their dynamics, at least for supersymmetric theories. Two successful frameworks for realizing five-dimensional QFTs within string/M-theory are provided by the worldvolume theories of $(p,q)$ 5-brane webs \cite{Aharony:1997bh} and by geometric engineering, where the theory arises from M-theory on an appropriate Calabi-Yau (CY) 3-fold conical singularity $X$ \cite{Morrison:1996xf}. These two approaches are not unrelated. Indeed, when $X$ is a toric CY 3-fold, the $(p,q)$-web corresponds to the toric spine dual to the toric diagram \cite{Leung:1997tw}.

As $(p,q)$ 5-branes can supersymmetrically end on $[p,q]$ 7-branes, it is natural to regard the $(p,q)$-web as hanging from 7-branes. Then, the space of mass deformations of the corresponding $5d$ QFT maps to the possible motions of the 7-branes. When all external legs of the $(p,q)$ web terminate on separate 7-branes, the resulting configuration is unconstrained. The situation becomes more subtle in the presence of parallel external legs, as several such legs may end on a common 7-brane. Different choices of 7-brane terminations then lead to distinct configurations and introduce additional constraints on the supersymmetry of the configuration through the so-called $s$-rule \cite{Hanany:1996ie,Bachas:1997sc,Kitao:1998mf,Mikhailov:1998bx,Benini:2009gi,Bergman:2020myx}. In \cite{Benini:2009gi}, a graphical approach for implementing these constraints was put forward. In short, $n$ parallel 5-brane legs correspond to $n$ parallel segments on the boundary of the toric diagram separated by $n-1$ dots. The pattern of 7-brane terminations can then be encoded by coloring in white the dots that separate external segments corresponding to legs ending on the same 7-brane, while keeping all other dots black. This results in a novel type of polytope dubbed {\it Generalized Toric Polygon} (GTP), which encodes generic $(p,q)$-webs suspended from 7-branes. In the special case in which all external legs terminate on distinct 7-branes, no white dots are present and the GTP reduces to an ordinary toric diagram. Consequently, the corresponding five-dimensional theory can be geometrically engineered by compactifying M-theory on the CY$_3$ singularity whose toric diagram is graph-dual to the associated $(p,q)$-web. Extending this correspondence to generic GTPs remains an open challenge and has recently attracted substantial attention (see \textit{e.g.} \cite{vanBeest:2020kou,VanBeest:2020kxw,Bourget:2023wlb,Arias-Tamargo:2024fjt,Alexeev:2024bko,Collinucci:2026kom} for a non-exhaustive list of references).

Very recently, it was realized in \cite{Arias-Tamargo:2024fjt,CarrenoBolla:2024fxy} that certain toric singularities $\mathbb{C}^3/\mathbb{Z}_{n^2}$, which admit a well-known desingularization into smoothed $T$-cones \cite{Shepherd1988}, play for GTPs the same role that elementary triangles play in triangulations of ordinary toric diagrams (see \cite{CarrenoBolla:2024fxy} for details). A smoothed $T$-cone (often simply called a $T$-cone) corresponds to an external leg configuration in which $n$ 5-branes terminate on a common 7-brane. In this language, a Hanany-Witten (HW) transition, obtained by moving a 7-brane across the $(p,q)$-web along its prong, corresponds to an algebraic mutation of the associated Laurent polynomial and is represented geometrically by flipping the corresponding $T$-cone from one side of the GTP to the other \cite{Arias-Tamargo:2024fjt}.

Importantly, this observation has two remarkable consequences. First, it endows the choice of origin of the polytope with a precise geometric meaning. Indeed, a HW transition flips the $T$-cone associated with the crossed 7-brane about its apex, naturally singling out that point as the origin of the polygon. In this context, it was shown in \cite{2012arXiv1212.1785A} that a suitably graded Hilbert series (HS), which coincides with the Ehrhart series of the dual polytope, is invariant under mutations generated by flipping a $T$-cone whose apex is located at the chosen origin. This new and more refined perspective is in contrast with the more traditional view of toric diagrams, which are often regarded up to translations, namely without assigning any special meaning to the origin. Second, since HW transitions often relate a standard toric diagram to a GTP, they suggest that the CY$_3$ geometry associated with a GTP should be closely related to the naive toric CY$_3$ obtained by simply replacing all white dots with black ones. More precisely, the two geometries should differ only by the freezing of a specific set of deformations as shown in \cite{Arias-Tamargo:2024fjt}. Taken together, these observations led \cite{Arias-Tamargo:2024fjt} to conjecture that the mutation-invariant HS, which reproduces the Ehrhart series of the dual polytope, can be obtained directly from the HS of the toric gauge theory associated with the underlying blackened toric diagram, provided the fields are assigned appropriate scalings. While several non-trivial examples supporting this conjecture were presented in \cite{Arias-Tamargo:2024fjt}, a systematic procedure for determining the required field scalings was lacking. The main goal of this paper is to fill this gap by providing an explicit algorithm for constructing the appropriate scaling assignments and computing the corresponding mutation-invariant Hilbert series.

As discussed above, our starting point is the toric gauge theory describing the fractional-brane quiver associated with the toric CY$3$ singularity obtained by treating the GTP as an ordinary toric diagram. As is well known, such a quiver gauge theory can be encoded in a {\it brane tiling}, also commonly referred to as a dimer model \cite{Hanany:2005ve,Franco:2005rj,Franco:2005sm}. When studying brane tilings, a central role is played by {\it perfect matchings}, which are collections of edges in the brane tiling such that every node is touched exactly once. Perfect matchings are conveniently encoded in the matrix $P_{e\alpha}$, whose entries are equal to 1 whenever the edge $e$ belongs to the perfect matching $p_{\alpha}$ and vanish otherwise. Perfect matchings can in turn be associated with points of the toric diagram. In general, except at the corners of the toric diagram, several perfect matchings are assigned to a given point. It turns out that the field scalings required for the gauge-theoretic Hilbert series to reproduce the geometric, mutation-invariant Hilbert series, which coincides with the Ehrhart series of the dual polytope for a chosen origin $\mathcal{O}$, take a remarkably simple form. We shall refer to these assignments as the \emph{Ehrhart scaling}. They are given by
\begin{equation}
\label{result}
\Delta(X_e)=\sum_{I \in \mathcal{P}_\mathcal{O}} P_{eI}\,,
\end{equation}
where $\mathcal{P}_\mathcal{O}$ is the set of perfect matchings at the chosen origin $\mathcal{O}$. In other words, the scaling of any field is given by the number of perfect matchings for the point at the origin that contain it. It should be emphasized that our proof relies on the assumption that every field in the gauge theory appears in at least one of the perfect matchings belonging to the set $\mathcal{P}_\mathcal{O}$. Although we do not currently have a general proof of this statement, we have verified its validity in a large number of non-trivial examples, several of which are presented below. Establishing this property rigorously would be very interesting and would further strengthen our construction.

The prescription in \eqref{result} is remarkably simple, suggesting that it may admit a deeper geometrical interpretation. It would be highly desirable to uncover the geometric principle underlying this construction. Similarly, we do not yet have a clear physical understanding of our prescription. It would therefore be very interesting to identify the physical rationale behind this particular choice of field scalings.

The recipe in \eqref{result} places on firm footing the conjecture put forward in \cite{Arias-Tamargo:2024fjt} that the brane tiling associated with the toric quiver gauge theory obtained by treating the GTP as an ordinary toric diagram retains non-trivial information about the GTP itself. Having said that, a much more ambitious goal is to identify a generalization of brane tilings for GTPs. We will revisit this question in a companion paper \cite{Franco:toappear}. It would be interesting to connect the scaling prescription introduced in this work to the construction in \cite{Franco:toappear}. 

The rest of this paper is organized as follows. In Section \sref{sec:proof}, we present a proof of the prescription in \eqref{result}. In Section \sref{Examples}, we validate our prescription through several examples. We begin by studying examples with a single internal point, for which the choice of origin is unique. We then turn to examples with two internal points. In these cases, different choices of origin lead to different GTPs (both toric diagrams and genuine GTPs with white points) and, consequently, to distinct $T$-cone tessellations. In the examples in Sections \sref{E1},\sref{E2} and \sref{C3/Z2xZ3}, we recover the scalings already inferred by trial-and-error in \cite{Arias-Tamargo:2024fjt}. In Section \sref{section_L251}, we study a new example and show agreement between our prescription and the geometric Hilbert series computed as the Ehrhart series of the dual polytope.

\section{From the Origin of the Toric Diagram to Field Scalings}
\label{sec:proof}

In this section, we prove the scaling prescription proposed in \eqref{result}. As is well known, the HS associated with a toric gauge theory, more precisely the HS of the coherent component of its moduli space \cite{Forcella:2008bb}, can be computed using standard field-theoretic techniques \cite{Benvenuti:2006qr} once a scaling assignment for the quiver fields has been specified. Among the many possible choices of scaling (see \cite{Bao:2024nyu} for a recent discussion), our goal is to identify the unique assignment for which the HS computed from the gauge theory reproduces the geometric HS given by the Ehrhart series of the dual polytope \cite{Arias-Tamargo:2024fjt}.\footnote{For brevity, we will often refer to the geometric HS computed as the Ehrhart series of the dual polytope simply as the Ehrhart series.} Let us be more precise about this point. The moduli space of a toric gauge theory is a toric CY$_3$ whose geometry is encoded in a convex lattice polygon $\Delta \subset \mathbb{Z}^2$, commonly referred to as the toric diagram. The HS we aim to reproduce coincides with the Ehrhart series of the dual polytope
\begin{equation}
\label{HSEhrhart}
HS=\sum_{n=0}|n\, \Delta^{\circ}\cap \mathbb{Z}^2|\,t^n\,,
\end{equation}
where the dual polytope $\Delta^{\circ}$ to the original one $\Delta$ is defined as 
\begin{equation}
\Delta^{\circ}=\{u\in \mathbb{Q}^2/\,\langle u| v\rangle \geq -1,\,\forall v\in \Delta\}\,.
\end{equation}
From \eqref{HSEhrhart}, we see that the HS counts, at degree $n$, the number of integral points within the $n$-times enlarged dual polytope. Our goal is to find a scaling for the fields in the toric gauge theory such that the HS of its coherent component matches \eqref{HSEhrhart}.

\subsection{Brane Tilings, Perfect Matchings and the Forward Algorithm}

Toric gauge theories can be fully encoded by brane tilings, which are represented by bipartite graphs on a 2-dimensional torus \cite{Hanany:2005ve,Franco:2005rj,Franco:2005sm} (see \textit{e.g.} \cite{Kennaway:2007tq,Yamazaki:2008bt} for reviews). Brane tilings significantly simplify the connection between the gauge theory and the underlying geometry in both directions. Faces $f_I$ of the brane tiling correspond to the gauge nodes of the quiver. Edges $e_{IJ}$ correspond to bifundamental fields between the gauge groups $f_I$ and $f_J$ (or adjoints, if $I=J$). Their orientation is determined by the bipartite structure of the graph: assigning, for example, a clockwise orientation to black nodes and a counterclockwise orientation to white nodes allows each edge to be regarded as an oriented edge $e_{IJ}$. Finally, every node of the graph corresponds to a monomial in the superpotential, given by the product of all the fields associated to the edges terminating on it. The sign of each term is conventionally fixed by the color of the corresponding node: black nodes contribute with a positive sign, while white nodes contribute with a negative sign.

Perfect matchings play a central role in the study of dimer models. A perfect matching $p_\alpha$ is a subset of edges such that every node in the graph $V$ is incident to exactly one edge in $p_\alpha$. The perfect matchings can be encoded in the perfect matching matrix $P$, whose entries indicate whether a given edge $e$ belongs to a perfect matching $p_\alpha$:
\begin{equation}
P_{e\alpha}=
\begin{cases}
1 & \text{if } e\in p_\alpha\,,\\
0 & \text{otherwise}\,.
\end{cases}
\end{equation}
By construction, $P$ is a $\#\,$fields $\times$ $\#\,$perfect matchings matrix.

As mentioned earlier, the edges of the brane tiling correspond to chiral fields in the gauge theory. It is useful to express chiral fields in terms of variables associated to perfect matchings as follows:
\begin{equation}
\label{Xinps}
X_e=\prod_{\alpha} p_{\alpha}^{P_{e\alpha}}\,.
\end{equation}
This map ensures that the $X_e$ automatically satisfy the vanishing $F$-term constraints \cite{Franco:2005rj}. As a result, perfect matchings are the natural variables in which to write a gauge linear sigma model whose target is the CY$_3$, which is is the modul space of the gauge theory. The procedure for determining the toric diagram of a CY$_3$ singularity from the corresponding gauge theory is known as the {\it forward algorithm} \cite{Feng:2000mi}. When reformulated in terms of the perfect matching parametrization, the algorithm becomes considerably more efficient and is commonly referred to as the {\it fast forward algorithm} \cite{Franco:2005rj}. One can encode the relations coming from $F$-terms into some charges for the perfect matchings according to $Q_F={\rm ker}\,P$. In addition, in order to impose the constraints coming from the vanishing of $D$-terms, one assigns gauge charges to the perfect matchings compatible with the gauge charges of the fields in the quiver via the map \eqref{Xinps}, constructing a matrix $Q_D$. One then appends $Q_F$ and $Q_D$ into
\begin{equation}
Q=\left(\begin{array}{c} Q_F \\ Q_D\end{array}\right)\,,
\end{equation}
and computes $G={\rm ker}\,Q$. The matrix $G$ is a 3$ \times \#\,$perfect matchings, whose columns correspond to the coordinates in $\mathbb{Z}^3$ of each perfect matching. The CY$_3$ condition ensures that all these points lie in a plane at height 1, and thus it is enough to consider the projection to the first $\mathbb{Z}^2$, where we find a polytope known as the toric diagram. In general, multiple perfect matchings are mapped to the same point in the toric diagram.

\subsection{The Field Scalings Leading to the Ehrhart Series}

Suppose now that we assign a scaling $\Delta_e$ to each field $X_e$. Assuming that scaling dimensions are additive, it follows from \eqref{Xinps} that
\begin{equation}
\Delta_e=\sum_{\alpha} w_{\alpha}\,P_{e\alpha}\,,
\end{equation}
where $w_{\alpha}$ denotes the scaling of the perfect matching $p_{\alpha}$. This parametrization automatically guarantees the homogeneity of the superpotential, i.e. that all monomials appearing in $W$ have the same scaling.\footnote{In fact, since, as mentioned below, every term in the superpotential is equal to the product of all perfect matchings, any scaling based on scaling assignments for individual perfect matchings leads to a homogeneous superpotential.} Indeed, $W$ can be written as a sum of monomials associated to the black and white vertices $V$ of the bipartite graph,
\begin{equation}
W=\sum_V W_V\,,
\qquad
W_V=\prod_{e\in V} X_e\,,
\end{equation}
and therefore
\begin{equation}
\label{DeltaW}
\Delta(W_V)=\sum_{e\in V} \Delta_e=\sum_{\alpha} w_{\alpha} \sum_{e\in V} P_{e\alpha} = \sum_{\alpha} w_{\alpha}\,,
\end{equation}
where we have used  the fact that each perfect matching touches every node exactly once. This is equivalent to the well-known fact that the definition of perfect matchings, together with the map between chiral fields and perfect matchings in \eqref{Xinps}, implies that every term in the superpotential is given by the product of all perfect matchings. This confirms that $\Delta(W_V)$ is independent of $V$, and therefore the superpotential $W$ is homogeneous.

We now want to determine a choice of the scalings $w_{\alpha}$ for each perfect matching such that the resulting field-theoretic HS matches \eqref{HSEhrhart}. Since all nodes of the quiver are assumed to have gauge group $U(1)$, the gauge-invariant operators counted by the HS are mesons. The field-theoretic HS is therefore a generating function whose coefficient at degree $n$ counts the number of mesonic operators with scaling $n$. Generically, we can write a meson in terms of the chiral fields as $M=\prod_e\,(X_e)^{m_e(M)}$, where $m_e(M)$ keeps track of the number of times the field $X_e$ appears in the meson $M$. In terms of perfect matchings, we have
\begin{equation}
M=\prod_{\alpha} p_{\alpha}^{n_{\alpha}(M)}\,,\qquad n_{\alpha}=\sum_em_e(M)\,P_{e\alpha}\,.
\end{equation}
The number $n_\alpha$ that a perfect matching $p_\alpha$ appears in a meson is obviously a non-negative integer. In the previous expression, this follows from the fact the entries in $P$ and $m_e(M)$ are positive semidefinite integers. The scaling of a meson becomes
\begin{equation}
\label{Deltameson}
\Delta(M)=\sum_{\alpha}\,n_{\alpha}(M)\,w_{\alpha}\,.
\end{equation}

We would like to map mesons to the dual lattice to connect to the definition of the Ehrhart series. To do that, consider the differences $n_{\alpha}(M)-n_{\beta}(M)$, which are given by
\begin{equation}
n_{\alpha}(M)-n_{\beta}(M)=\sum_e m_e(M)\,(P_{e\alpha}-P_{e\beta})\,.
\end{equation}
The difference of two perfect matchings defines a closed 1-cycle on the torus, which is related to $P_{e\alpha}-P_{e\beta}$ in the previous expression. Thus, we can reinterpret this difference as
\begin{equation}
n_{\alpha}(M)-n_{\beta}(M)=\langle m(M) | p_{\alpha}-p_{\beta}\rangle\,,
\end{equation}
where the kets refer to elements of $H^1(\mathbb{T}^2)$ and bras to the dual space. Kets correspond to closed cycles in the brane tiling, while bras correspond to closed loops in the dual periodic quiver representing mesons. This formula reflects the known correspondence between mesons and elements of the dual lattice.

As reviewed earlier, the fast forward algorithm maps perfect matchings to points in the toric diagram. Equivalently, a point in the toric diagram can be associated with the difference between a given perfect matching and a reference perfect matching. In this picture, the position of the corresponding point is determined by the slope of the resulting height function, or equivalently by the monodromies of the height function along the two fundamental cycles of $\mathbb{T}^2$. In general, this correspondence is many-to-one: multiple perfect matchings can map to the same point in the toric diagram. Let us focus on an internal point $\mathcal{O}$ of the toric diagram. Let $\mathcal{P}_{\mathcal O}$ denote the set of perfect matchings associated with $\mathcal O$, which we label as $p_I$, with $I=1,\ldots, |\mathcal{P}_{\mathcal O}|$. Choosing $\mathcal{O}$ as the origin corresponds to selecting a perfect matching $p_I \in \mathcal{P}_{\mathcal O}$ as the reference. Then, $|v_{\alpha}\rangle =|p_{\alpha}-p_I\rangle$ stands for the coordinates of the corresponding point in the toric diagram. Obviously, with this choice of reference, we obtain $|v_{J}\rangle=(0,0)^T$. In this notation, 
\begin{equation}
n_{\alpha}(M)-n_I(M)=\langle m(M) | v_{\alpha}\rangle\,.
\label{nalpha_minus_nO}
\end{equation}
Notice that if we choose $p_\alpha$ to be any other perfect matching associated with the origin, namely $p_\alpha = p_J$, then $n_I= n_J$. In other words, all perfect matchings associated with the origin $\mathcal O$ appear with the same multiplicity in any given meson.\footnote{In fact this property is true for all perfect matchings corresponding to the same point in the toric diagram, regardless of whether that point is the origin. This can be shown by the straightforward generalization of \eqref{nalpha_minus_nO}.}

Moreover, since $n_{\alpha}(M)\geq 0$, we have
\begin{equation}
\langle m(M) | v_{\alpha}\rangle\geq -n_{I}(M)\,.
\end{equation}
We see that $n$, the dimension of a meson in \eqref{HSEhrhart} precisely corresponds to $n_I(M)$. Recall that everything is independent of the chosen $p_I$ since, as we said, $n_I(M)=n_J(M)$ for all $p_I,p_j\in \mathcal{P}_\mathcal{O}$. For reasons which will be clear momentarily, we sum over all the (equivalent) $I$, and define the Ehrhart scaling as 
\begin{equation}
\Delta(M)=\sum_{I \in \mathcal{P}_\mathcal{O}}\,n_I(M)\,.
\end{equation}
Comparing this last formula to \eqref{Deltameson}, we see that the $w_{\alpha}$ are
\begin{equation}
w_{\alpha}=
\begin{cases}
1 & \text{if } \alpha\in \mathcal{P}_\mathcal{O} \,,\\
0 & \text{otherwise}\,.
\end{cases}
\end{equation}
The scaling of each chiral field is therefore given by
\begin{equation}
\Delta_e=\sum_{I \in  \mathcal{P}_\mathcal{O}} P_{eI}\, .
\end{equation}
That is, the scaling of the field $X_e$ is simply given by the sum of the entries in the row of the $P$-matrix corresponding to $X_e$, restricted to the perfect matchings in $\mathcal{P}_{\mathcal O}$. Simply put, it is the number of perfect matchings associated with the origin that contain $X_e$. This is the prescription anticipated in \eqref{result}. Together with \eqref{DeltaW}, it implies that the superpotential is a homogeneous function of degree equal to $|\mathcal{P}_\mathcal{O}|$, that is, the number of perfect matchings at the chosen origin.

There remains one loose end to fully complete our argument. One might worry about the possibility that some field $X_e$ does not appear in any of the perfect matchings in the set $\mathcal{P}_{\mathcal O}$, i.e. that it is not contained in any of the perfect matchings associated with the origin. In such a situation, our prescription would assign a vanishing scaling to $X_e$, which would be inadmissible.

Inspection of a large number of explicit examples indicates that every field in the theory indeed appears in at least one perfect matching belonging to $\mathcal{P}_{\mathcal O}$ (for any possible choice of $\mathcal{O}$, i.e. for any internal point). It would be extremely desirable to fill this gap and determine whether this is indeed the case. In fact, this observation motivated the sum over all perfect matchings in $\mathcal{P}_{\mathcal O}$ in our prescription. Had we selected a single perfect matching in $\mathcal{P}_{\mathcal O}$ rather than summing over all of them, some fields would generally fail to appear in the chosen perfect matching, leading to an inconsistent scaling assignment. A similar issue would arise if we chose a boundary point as the origin.\footnote{Boundary points can also be ruled out as possible origins on independent grounds, since they could not serve as the apex of a consistent tessellation of a GTP.}

More generally, even if there existed internal points whose associated perfect matchings failed to contain some of the chiral fields, this need not invalidate our construction. Rather, it may indicate that such points cannot serve as the origin of a consistent GTP. This suggests an intriguing criterion for identifying admissible choices of origin directly from the perfect matching structure.

\section{Examples}\label{Examples}

We now consider various examples and confirm that our prescription gives rise to scalings with the desired properties.

\subsection{Toric Diagrams with One Internal Point}

Let us start with the simplest class of examples, namely those where the toric diagram contains only one internal point. In these cases, there is only one possible choice of origin and, consequently, a unique $T$-cone tessellation. Moreover, all these examples are ordinary toric diagrams and the $T$-cones are minimal triangles. Nevertheless, their behavior under polytope mutations is non-trivial.

\subsubsection{Complex Cone over $\mathbb{F}_2$}\label{E1}

Our first example is the complex cone over $\mathbb{F}_2$, whose associated $5d$ SCFT is the $E_1$ theory of \cite{Seiberg:1996bd}. This example is studied in Section 5.1 of \cite{Arias-Tamargo:2024fjt}.\footnote{For the examples discussed in Sections \sref{E1},\sref{E2} and \sref{C3/Z2xZ3}, the corresponding quivers, superpotentials and brane tilings were presented in \cite{Arias-Tamargo:2024fjt}. To avoid unnecessary repetition, we refer the reader to that reference for their explicit form. Throughout our analysis, we adopt the same labeling conventions for nodes and fields when determining perfect matchings.} The perfect matchings are
 \begin{equation}
\begin{array}{clcclccl}
p_1 &= \{X_{31},X_{41},Y_{41},X_{42}\} & \ \ & p_2 & = \{X_{13},X_{24},X_{31},X_{42}\}  &\ \ & p_3 & = \{X_{13},X_{23},Y_{23},X_{24}\} \\
p_4 &= \{X_{12},Y_{12},X_{34},Y_{34}\} & \ \ & p_5 & = \{Y_{12},Y_{23},X_{34},X_{41}\}  & \ \ & p_6 &= \{X_{24},X_{31},X_{34},Y_{34}\} \\
p_7 & = \{X_{12},Y_{12},X_{13},X_{42}\} & \ \ & p_8 & = \{X_{12},X_{23},Y_{34},Y_{41}\} & \ \  & p_9 & = \{X_{23},Y_{23},X_{41},Y_{41}\} 
\end{array}
\end{equation}

\begin{figure}[H]
\centering
\includegraphics[width=4cm]{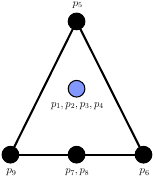}
\caption{Toric diagram for the complex cone over $\mathbb{F}_2$ and perfect matchings corresponding to each point. The origin is indicated in blue.}
\label{toric_F2}
\end{figure}

\fref{toric_F2} shows the $\mathbb{F}_2$ toric diagram and the perfect matchings corresponding to each point.

From \fref{toric_F2}, we see that $\mathcal{P}_\mathcal{O} = \{p_1,p_2,p_3,p_4\}$. The restriction of the $P$-matrix to the perfect matchings at the origin is
\begin{equation}
P|_{\mathcal{P}_\mathcal{O}}=\left( \begin{array}{c | c c c c c c c c c c c c} 
& X_{12} & Y_{12} & X_{13} & X_{23} & Y_{23} & X_{24} & X_{31} & X_{34} & Y_{34} & X_{41} & Y_{41} & X_{42} \\ \hline
p_1 & 0 & 0 & 0 & 0 & 0 & 0 & 1  & 0 & 0 & 1 & 1 & 1 \\
p_2 & 0 & 0 & 1 & 0 & 0 & 1 & 1  & 0 & 0 & 0 & 0 & 1 \\
p_3 & 0 & 0 & 1 & 1 & 1 & 1 & 0  & 0 & 0 & 0 & 0 & 0 \\
p_4 & 1 & 1 & 0 & 0 & 0 & 0 & 0  & 1 & 1 & 0 & 0 & 0 
\end{array}\right)
\end{equation}
As anticipated, every field in the quiver appears at least once in this set. Our prescription then assigns to each field a scaling dimension given by the sum of the entries in the corresponding column, resulting in
\begin{equation}
\Delta(\{X_{12},Y_{12}, X_{23}, Y_{23} , X_{34} , Y_{34} , X_{41} , Y_{41}\})=1\,,\qquad \Delta(\{X_{24},X_{42},X_{13},X_{31}\})=2\,,
\end{equation}
which is precisely the scaling found in \cite{Arias-Tamargo:2024fjt}. Also note that $\Delta(W)=4$, which is indeed the number of perfect matchings at the origin.

\subsubsection{Complex Cone over $PdP_2$}\label{E2}

Let us now consider the complex cone over $PdP_2$. Its corresponding $5d$ SCFT is the $E_2$ theory of \cite{Seiberg:1996bd}. This example was studied in Section 6.1 of \cite{Arias-Tamargo:2024fjt}. The perfect matchings are
\begin{equation}
\begin{array}{clcclccl}
p_1 &= \{X_{41}, X_{42}, X_{51},  X_{52}\} & \ \ & p_2 &= \{X_{13}, X_{23}, Y_{23}, X_{24}\} & \ \ & p_3 &= \{X_{12}, X_{13}, X_{42}, X_{52}\} \\
p_4 &= \{X_{24}, X_{34}, Y_{34}, X_{35}\} & \ \ & p_5 &= \{X_{35}, X_{41}, X_{42}, X_{45} \} & \ \ & p_6 &= \{X_{13}, X_{24}, X_{35}, X_{42}\} \\
p_7 &= \{X_{23}, X_{24},Y_{34},X_{51}\} & \ \ & p_8 &= \{X_{12}, X_{23}, Y_{34}, X_{45}\} & \ \ & p_9 &= \{X_{12}, X_{34}, Y_{34}, X_{52}\} \\
p_{10} &= \{X_{23}, Y_{23}, X_{41}, X_{45}\} & \ \ & p_{11} &= \{Y_{23}, X_{34}, X_{41}, X_{52}\} & \ \ &
\end{array}
\end{equation}

\begin{figure}[H]
\centering
\includegraphics[width=4cm]{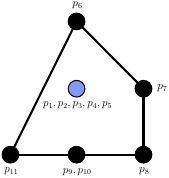}
\caption{Toric diagram for $PdP_2$ and perfect matchings corresponding to each point. The origin is indicated in blue.}
\label{toric_PdP2}
\end{figure}

\fref{toric_PdP2} shows the $\mathbb{F}_2$ toric diagram and the perfect matchings corresponding to each point.

In this case, $\mathcal{P}_\mathcal{O}=\{p_1,p_2,p_3,p_4,p_5\}$. Indeed, all fields in the quiver appear within this set. The $P$-matrix reduced to $\mathcal{P}_\mathcal{O}$ is

\begin{equation}
P^T|_{\mathcal{P}_\mathcal{O}}=
\left(
\begin{array}{c|ccccccccccccc}
 & X_{12} & X_{13} & X_{23} & Y_{23} & X_{24} & X_{34} & Y_{34} & X_{35} & X_{41} & X_{42} & X_{45} & X_{51} & X_{52} \\ \hline
p_1 & 0 & 0 & 0 & 0 & 0 & 0 & 0 & 0 & 1 & 1 & 0 & 1 & 1 \\
p_2 & 0 & 1 & 1 & 1 & 1 & 0 & 0 & 0 & 0 & 0 & 0 & 0 & 0 \\
p_3 & 1 & 1 & 0 & 0 & 0 & 0 & 0 & 0 & 0 & 1 & 0 & 0 & 1 \\
p_4 & 0 & 0 & 0 & 0 & 1 & 1 & 1 & 1 & 0 & 0 & 0 & 0 & 0 \\
p_5 & 0 & 0 & 0 & 0 & 0 & 0 & 0 & 1 & 1 & 1 & 1 & 0 & 0 
\end{array}
\right)
\end{equation}
Then, 
\begin{equation}
\begin{array}{l}
\Delta(X_{42}) = 3\,, \\[.1cm]
\Delta(\{X_{13},X_{24},X_{35},X_{41},X_{52}\})= 2\,, \\[.1cm] 
\Delta(\{X_{12},X_{23},Y_{23},X_{34},Y_{34},X_{45},X_{51}\}) = 1\,,
\end{array}
\end{equation}
which is the scaling found in \cite{Arias-Tamargo:2024fjt}. As a simple check, note that $\Delta(W)=5$, the number of perfect matchings at the origin.

\subsection{Examples with Two Internal Points}

Let us now consider examples with two internal points. These cases are interesting, since there is freedom in the choice of the origin. Each choice corresponds to a different $T$-cone tessellation and, equivalently, to a different brane web and $5d$ theory.

\subsubsection{$\mathbb{C}^3/(\mathbb{Z}_2\times \mathbb{Z}_3)$}\label{C3/Z2xZ3}

\begin{figure}[H]
\centering
\includegraphics[width=7.7cm]{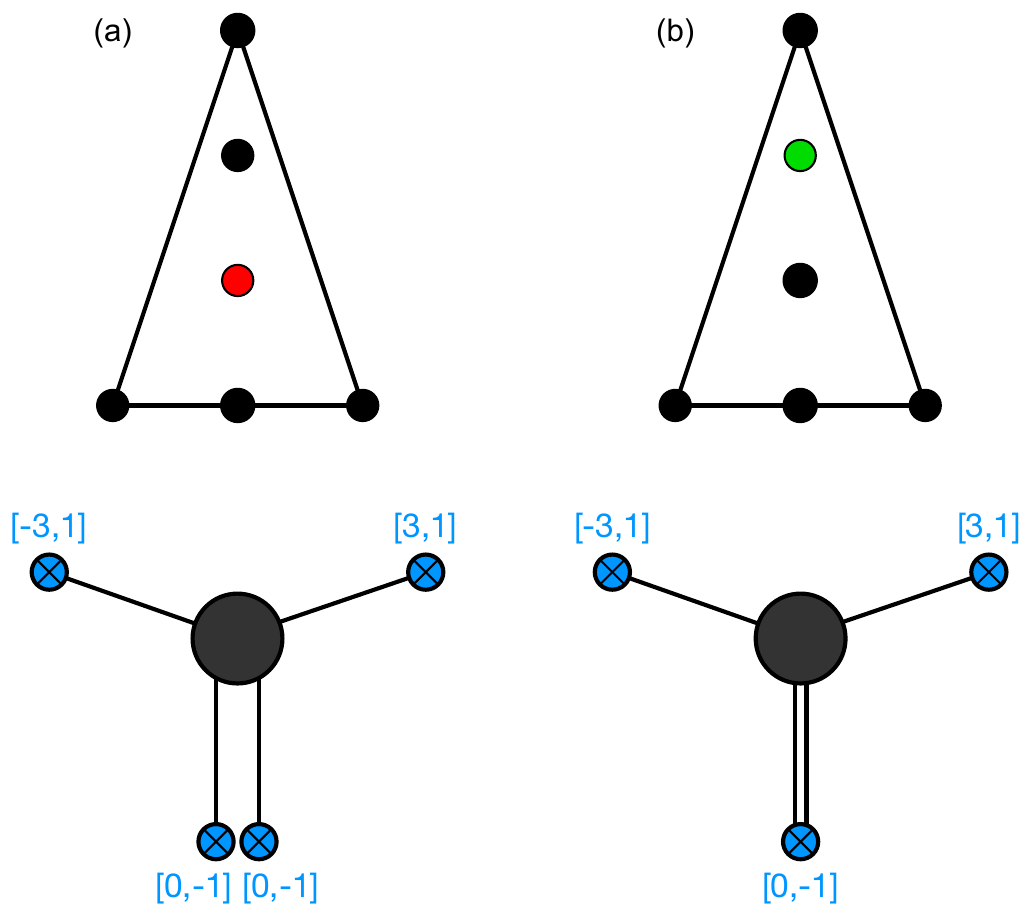}
\caption{Toric diagram for a $\mathbb{C}^3/(\mathbb{Z}_2\times \mathbb{Z}_3)$ orbifold and two possible choices of the origin. a) Choice corresponding to a web where every leg terminates on a different 7-brane. b) Choice corresponding to a web with two legs ending on the same 7-brane.}
\label{Z2xZ3}
\end{figure}

Consider the polytope in \fref{Z2xZ3}. As a toric diagram, it corresponds to $\mathbb{C}^3/(\mathbb{Z}_2\times \mathbb{Z}_3)$ with orbifold weights
$(1,-1,0)$ and $(1,1,1)$. There are two possible choices of the origin, each of them corresponding to a different brane web, as shown in \fref{Z2xZ3}. For choice (a), every 5-brane leg terminates on an independent 7-brane, so the polytope is indeed an ordinary toric diagram. In contrast, for choice (b) the two vertical legs terminate on the same 7-brane, so the corresponding polytope should therefore be regarded as a GTP with white dots. Accordingly, depending on the choice of origin, the lower edge either subtends two $T_1$-cones (red) or a single $T_2$-cone (green).

\begin{figure}[H]
\centering
\includegraphics[width=4cm]{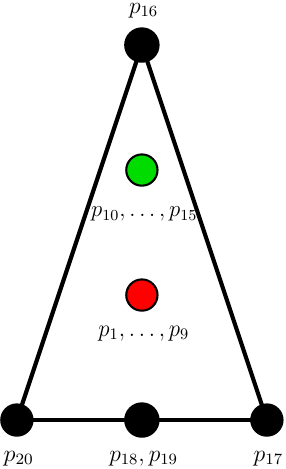}
\caption{Perfect matchings for each point in the $\mathbb{C}^3/(\mathbb{Z}_2\times \mathbb{Z}_3)$ toric diagram.}
\label{Z2xZ3_pms}
\end{figure}

This example is studied in Section 6.2 of \cite{Arias-Tamargo:2024fjt}. The perfect matchings are
\begin{equation}
\small
\begin{array}{clccl}
p_1 &= \{X_{21}, X_{26}, Y_{26}, X_{35}, X_{45},  Y_{45}\}  &&
p_1 &= \{X_{32}, Y_{32}, X_{35}, X_{42}, X_{45},  Y_{45}\}  \\
p_3 &= \{X_{13}, Y_{13}, X_{42}, X_{45}, Y_{45}, X_{63}\} &&
p_4 &= \{X_{13}, Y_{13}, X_{14}, X_{26}, Y_{26}, X_{56}\}  \\
p_5 &= \{X_{21}, X_{26},Y_{26}, X_{51},Y_{51}, X_{56}\}  &&
p_6 &= \{X_{32}, Y_{32}, X_{42}, X_{51}, Y_{51}, X_{56}\} \\
p_7 &= \{X_{14}, X_{32}, Y_{32}, X_{35}, X_{64}, Y_{64}\}  &&
p_8 &= \{X_{13}, Y_{13}, X_{14}, X_{63}, X_{64}, Y_{64}\} \\
p_9 &= \{X_{21}, X_{51}, Y_{51}, X_{63}, X_{64}, Y_{64}\} &&
p_{10} &= \{X_{21}, X_{35}, X_{42}, X_{45}, Y_{45}, X_{63}\} \\
p_{11} &= \{X_{14}, X_{21}, X_{26}, Y_{26}, X_{35}, X_{56}\} &&
p_{12} &= \{X_{14}, X_{32}, Y_{32}, X_{35}, X_{42}, X_{56}\}  \\
p_{13} &= \{X_{13}, Y_{13}, X_{14}, X_{42}, X_{56}, X_{63}\} &&
p_{14} &= \{X_{21}, X_{42}, X_{51}, Y_{51}, X_{56}, X_{63}\}  \\
p_{15} &= \{X_{14}, X_{21}, X_{35}, X_{63}, X_{64}, Y_{64}\} &&
p_{16} &= \{X_{14}, X_{21}, X_{35},  X_{42}, X_{56}, X_{63}\}  \\
p_{17} &= \{X_{13}, Y_{26}, X_{32}, Y_{45}, Y_{51}, X_{64}\}  &&
p_{18} &= \{X_{13}, Y_{13}, X_{26}, Y_{26}, X_{45}, Y_{45}\} \\
p_{19} &= \{X_{32},Y_{32}, X_{51}, Y_{51}, X_{64}, Y_{64}\}  &&
p_{20} &= \{Y_{13}, X_{26},Y_{32}, X_{45}, X_{51},  Y_{64}\}  
\end{array}
\end{equation}
Their positions in the toric diagram are shown in \fref{Z2xZ3_pms}.

Let us now consider how our prescription for scalings applies to each choice of origin.

\subsubsection*{Red Origin}

Let us first consider the red dot as the origin. In this case, $\mathcal{P}_{\mathcal{O}^{\rm red}}=\{p_1, \ldots, p_9 \}$. The $P$-matrix truncated to $\mathcal{P}_{\mathcal{O}^{\rm red}}$ becomes
{\footnotesize
\begin{equation}
P^T|_{\mathcal{P}_{\mathcal{O}^{\rm red}}}=
\left(
\begin{array}{c|cccccccccccccccccc}
 & X_{13} & Y_{13} & X_{14} & X_{21} & X_{26} & Y_{26} & X_{32} & Y_{32} & X_{35} & X_{42} & X_{45} & Y_{45} & X_{51} & Y_{51} & X_{56} & X_{63} & X_{64} & Y_{64} \\ \hline
p_1 & 0 & 0 & 0 & 1 & 1 & 1 & 0 & 0 & 1 & 0 & 1 & 1 & 0 & 0 & 0 & 0 & 0 & 0 \\
p_2 & 0 & 0 & 0 & 0 & 0 & 0 & 1 & 1 & 1 & 1 & 1 & 1 & 0 & 0 & 0 & 0 & 0 & 0 \\
p_3 & 1 & 1 & 0 & 0 & 0 & 0 & 0 & 0 & 0 & 1 & 1 & 1 & 0 & 0 & 0 & 1 & 0 & 0 \\
p_4 & 1 & 1 & 1 & 0 & 1 & 1 & 0 & 0 & 0 & 0 & 0 & 0 & 0 & 0 & 1 & 0 & 0 & 0 \\
p_5 & 0 & 0 & 0 & 1 & 1 & 1 & 0 & 0 & 0 & 0 & 0 & 0 & 1 & 1 & 1 & 0 & 0 & 0 \\
p_6 & 0 & 0 & 0 & 0 & 0 & 0 & 1 & 1 & 0 & 1 & 0 & 0 & 1 & 1 & 1 & 0 & 0 & 0 \\
p_7 & 0 & 0 & 1 & 0 & 0 & 0 & 1 & 1 & 1 & 0 & 0 & 0 & 0 & 0 & 0 & 0 & 1 & 1 \\
p_8 & 1 & 1 & 1 & 0 & 0 & 0 & 0 & 0 & 0 & 0 & 0 & 0 & 0 & 0 & 0 & 1 & 1 & 1 \\
p_9 & 0 & 0 & 0 & 1 & 0 & 0 & 0 & 0 & 0 & 0 & 0 & 0 & 1 & 1 & 0 & 1 & 1 & 1 
\end{array}
\right)
\end{equation}
}
Once again, all chiral fields are contained in the set $\mathcal{P}_{\mathcal{O}^{\rm red}}$. According to our prescription,
\begin{equation}
\Delta(\{ {\rm all\,fields}\})=3\,,
\end{equation}
reproducing the result in \cite{Arias-Tamargo:2024fjt}.\footnote{Strictly speaking, the scaling found in \cite{Arias-Tamargo:2024fjt} was $\Delta(\{ {\rm all\,fields}\})=1$. What really matters is that the scaling is the same for all fields.} As expected, $\Delta(W)=9$, which is the number of perfect matchings at the origin.

\subsubsection*{Green Origin}

For green origin, $\mathcal{P}_{\mathcal{O}^{\rm green}}=\{p_{10},\ldots,p_{15}\}$. The corresponding $P$-matrix is
\begin{equation}
\resizebox{1\textwidth}{!}{$
P^T|_{\mathcal{P}_{\mathcal{O}^{\rm green}}}=
\left(
\begin{array}{c|cccccccccccccccccc}
 & X_{13} & Y_{13} & X_{14} & X_{21} & X_{26} & Y_{26} & X_{32} & Y_{32} & X_{35} & X_{42} & X_{45} & Y_{45} & X_{51} & Y_{51} & X_{56} & X_{63} & X_{64} & Y_{64} \\ \hline
p_{10} & 0 & 0 & 0 & 1 & 0 & 0 & 0 & 0 & 1 & 1 & 1 & 1 & 0 & 0 & 0 & 1 & 0 & 0 \\
p_{11} & 0 & 0 & 1 & 1 & 1 & 1 & 0 & 0 & 1 & 0 & 0 & 0 & 0 & 0 & 1 & 0 & 0 & 0 \\
p_{12} & 0 & 0 & 1 & 0 & 0 & 0 & 1 & 1 & 1 & 1 & 0 & 0 & 0 & 0 & 1 & 0 & 0 & 0 \\
p_{13} & 1 & 1 & 1 & 0 & 0 & 0 & 0 & 0 & 0 & 1 & 0 & 0 & 0 & 0 & 1 & 1 & 0 & 0 \\
p_{14} & 0 & 0 & 0 & 1 & 0 & 0 & 0 & 0 & 0 & 1 & 0 & 0 & 1 & 1 & 1 & 1 & 0 & 0 \\
p_{15} & 0 & 0 & 1 & 1 & 0 & 0 & 0 & 0 & 1 & 0 & 0 & 0 & 0 & 0 & 0 & 1 & 1 & 1 
\end{array}
\right)
$}
\end{equation}
As one can check, all fields appear in $\mathcal{P}_{\mathcal{O}^{\rm green}}$. Using our procedure, the scalings become
\begin{align}
& \Delta(\{X_{14}, X_{21}, X_{35}, X_{42},X_{56}, X_{63}\})=4\,,\\
& \Delta(\{X_{13},Y_{13},X_{26},Y_{26},X_{32},Y_{32},X_{45},Y_{45},X_{51},Y_{51},X_{64},Y_{64}\})=1\,.\nonumber
\end{align} 
recovering the results in \cite{Arias-Tamargo:2024fjt}. Again, $\Delta(W)=6$, which is the number of perfect matchings at the origin.

\subsubsection{$L^{2,5,1}$}

\label{section_L251}

\begin{figure}[H]
\centering
\includegraphics[width=9.5cm]{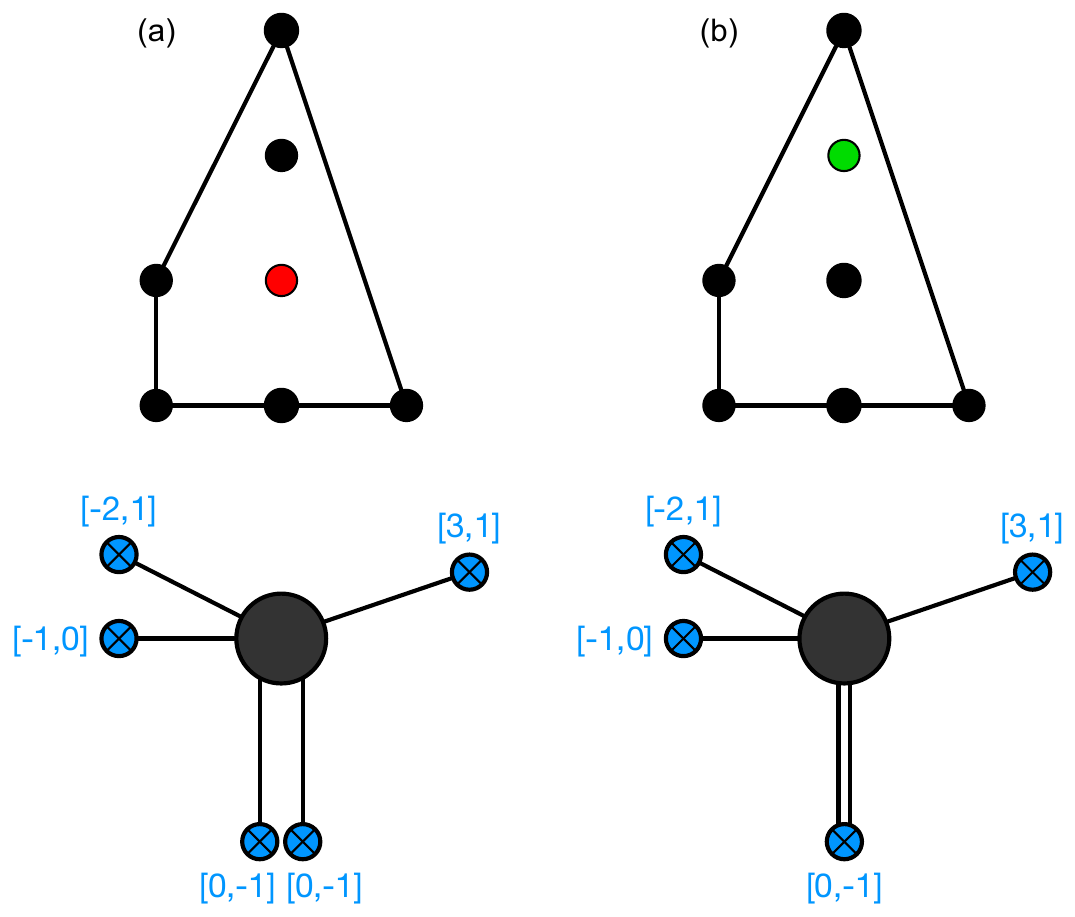}
\caption{Toric diagram for the cone over $L^{2,5,1}$ and two possible choices of
the origin. a) Choice corresponding to a web where every leg terminates on a different
7-brane. b) Choice corresponding to a web with two legs ending on the same 7-brane.}
\label{L251}
\end{figure}

Consider the polytope in \fref{L251}, which corresponds to $L^{2,5,1}$. The infinite family of 5-dimensional Sasaki-Einstein manifolds $L^{a,b,c}$ was introduced and studied in \cite{Cvetic:2005ft,Martelli:2005wy}. The corresponding CY 3-fold cones define an infinite class of toric singularities, and the associated gauge theories and brane tilings were constructed in \cite{Franco:2005sm,Butti:2005sw,Benvenuti:2005ja}. This theory has not been previously studied from the viewpoint of scalings and the Ehrhart series in \cite{Arias-Tamargo:2024fjt}. However, it has been recently considered in \cite{Kho:2026zwc} from the perspective of GTPs arising from polytope mutations and the corresponding cluster integrable systems. Once again, there are two possible choices of the origin and each of them corresponds to a different brane web as shown in \fref{L251}. For choice (a), every 5-brane leg terminates on an independent 7-brane, so the polytope is an ordinary toric diagram. For choice (b), the two vertical legs terminate on the same 7-brane, so the corresponding polytope should be interpreted as a GTP with white dots. As in the example in the previous section, depending on the origin, the lower side is the base of two $T_1$-cones (red) or a single $T_2$-cone (green).

The brane tiling for this theory was recently presented in \cite{Kho:2026zwc}, whose notation we follow. From it, we can read off the quiver, which has 7 nodes and the following superpotentials
\begin{align}
W=& X_{17}X_{75}X_{51}-X_{75}X_{54}X_{47}+X_{54}X_{43}X_{35}-X_{43}X_{32}X_{24}+X_{32}X_{21}X_{13}-X_{21}X_{17}X_{76}X_{62}\nonumber \\
& - X_{16}X_{65}X_{51}+X_{65}Y_{54}X_{47}X_{76}-Y_{54}Y_{43}X_{35}+Y_{43}Y_{32}X_{24}-Y_{32}Y_{21}X_{13}+Y_{21}X_{16}X_{62}\,.\end{align}

\begin{figure}[H]
\centering
\includegraphics[width=4cm]{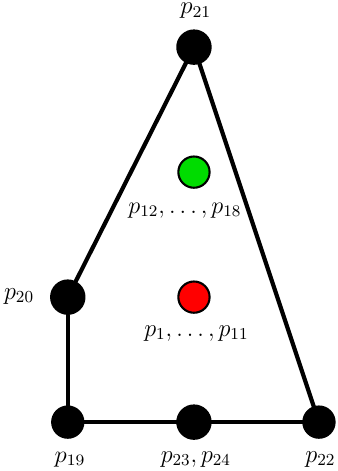}
\caption{Perfect matchings for each point in the toric diagram for the cone over $L^{2,5,1}$.}
\label{L251_pms}
\end{figure}

The perfect matchings are 
\begin{equation}
\small
\begin{array}{clccl}
p_1 &= \{X_{16}, X_{17}, X_{32}, Y_{32}, X_{35}, X_{47}\} &&
p_2 &= \{X_{13}, X_{16}, X_{17}, X_{43}, Y_{43}, X_{47}\} \\
p_3 &= \{X_{13}, X_{16}, X_{17}, X_{24}, X_{54}, Y_{54}\}&&
p_4 &= \{X_{21}, Y_{21}, X_{43}, Y_{43}, X_{47}, X_{51}\} \\
p_5 &= \{X_{21}, Y_{21}, X_{24}, X_{51}, X_{54}, Y_{54}\} &&
p_6 &= \{X_{32}, Y_{32},  X_{51}, X_{54}, Y_{54}, X_{62}\} \\
p_7 &= \{X_{21}, Y_{21}, X_{24}, X_{35}, X_{65}, X_{75}\} &&
p_8 &= \{X_{16}, X_{32}, Y_{32}, X_{35}, X_{75},  X_{76}\} \\
p_9 &= \{X_{32}, Y_{32}, X_{35}, X_{62}, X_{65}, X_{75}\} &&
p_{10} &= \{X_{13}, X_{16}, X_{43}, Y_{43}, X_{75},   X_{76}\} \\
p_{11} &= \{X_{13}, X_{43}, Y_{43}, X_{62}, X_{65}, X_{75}\} &&
p_{12} &= \{ X_{13}, X_{16}, X_{17}, X_{24},X_{35},X_{47}\} \\
p_{13} &= \{X_{21}, Y_{21}, X_{24}, X_{35}, X_{47}, X_{51}\} &&
p_{14} &= \{X_{32}, Y_{32}, X_{35}, X_{47}, X_{51},  X_{62}\} \\
p_{15} &= \{X_{13}, X_{43}, Y_{43}, X_{47}, X_{51}, X_{62}\} &&
p_{16} &= \{X_{13}, X_{24}, X_{51}, X_{54}, Y_{54}, X_{62}\} \\
p_{17} &= \{X_{13}, X_{16}, X_{24}, X_{35}, X_{75},  X_{76}\} &&
p_{18} &= \{X_{13}, X_{24}, X_{35}, X_{62}, X_{65}  X_{75}\} \\
p_{19} &= \{X_{17}, Y_{21}, X_{32}, Y_{43}, X_{54},  X_{65} \} &&
p_{20} &= \{Y_{21}, X_{32}, Y_{43}, X_{51}, X_{54},  X_{76}\} \\
p_{21} &= \{X_{13}, X_{24}, X_{35}, X_{47}, X_{51}, X_{62}\} &&
p_{22} &= \{X_{16}, X_{21},Y_{32}, X_{43}, Y_{54}, X_{75}\} \\
p_{23} &= \{X_{17}, X_{16}, X_{32}, Y_{32}, X_{54}, Y_{54}\} &&
p_{24} &= \{X_{21}, Y_{21}, X_{43}, Y_{43}, X_{65}, X_{75}\}
\end{array}
\end{equation} 
\fref{L251_pms} shows the positions of the perfect matchings in the toric diagram.

Let us now consider the scalings associated with each choice of origin.

\subsubsection*{Red Origin}

Considering the red origin, we have $\mathcal{P}_{\mathcal{O}^{\rm red}}=\{p_1,\ldots,p_{11}\}$ and
\begin{equation}
\resizebox{1\textwidth}{!}{$
P^T|_{\mathcal{P}_{\mathcal{O}^{\rm red}}}=
\left(
\begin{array}{c|ccccccccccccccccccc}
 & X_{13} & X_{16} & X_{17} & X_{21} & Y_{21}  & X_{24} & X_{32} & Y_{32} & X_{35} & X_{43} & Y_{43} & X_{47} & X_{51} & X_{54} & Y_{54} & X_{62} & X_{65} & X_{75} & X_{76} \\ \hline
p_1  & 0 & 1 & 1 & 0 & 0 & 0 & 1 & 1 & 1 & 0 & 0 & 1 & 0 & 0 & 0 & 0 & 0 & 0 & 0 \\
p_2  & 1 & 1 & 1 & 0 & 0 & 0 & 0 & 0 & 0 & 1 & 1 & 1 & 0 & 0 & 0 & 0 & 0 & 0 & 0 \\
p_3  & 1 & 1 & 1 & 0 & 0 & 1 & 0 & 0 & 0 & 0 & 0 & 0 & 0 & 1 & 1 & 0 & 0 & 0 & 0 \\
p_4 & 0 & 0 & 0 & 1 & 1 & 0 & 0 & 0 & 0 & 1 & 1 & 1 & 1 & 0 & 0 & 0 & 0 & 0 & 0 \\
p_5 & 0 & 0 & 0 & 1 & 1 & 1 & 0 & 0 & 0 & 0 & 0 & 0 & 1 & 1 & 1 & 0 & 0 & 0 & 0 \\
p_6 & 0 & 0 & 0 & 0 & 0 & 0 & 1 & 1 & 0 & 0 & 0 & 0 & 1 & 1 & 1 & 1 & 0 & 0 & 0 \\
p_7 & 0 & 0 & 0 & 1 & 1 & 1 & 0 & 0 & 1 & 0 & 0 & 0 & 0 & 0 & 0 & 0 & 1 & 1 & 0 \\
p_6 & 0 & 1 & 0 & 0 & 0 & 0 & 1 & 1 & 1 & 0 & 0 & 0 & 0 & 0 & 0 & 0 & 0 & 1 & 1 \\
p_9 & 0 & 0 & 0 & 0 & 0 & 0 & 1 & 1 & 1 & 0 & 0 & 0 & 0 & 0 & 0 & 1 & 1 & 1 & 0 \\
p_{10} & 1 & 1 & 0 & 0 & 0 & 0 & 0 & 0 & 0 & 1 & 1 & 0 & 0 & 0 & 0 & 0 & 0 & 1 & 1 \\
p_{11} & 1 & 0 & 0 & 0 & 0 & 0 & 0 & 0 & 0 & 1 & 1 & 0 & 0 & 0 & 0 & 1 & 1 & 1 & 0 
\end{array}
\right)
$}
\end{equation}

Then, according to our prescription, the scalings are
\begin{equation}
\begin{array}{l}
\Delta(\{X_{76}\})=2\,, \\[.1cm]
\Delta(\{X_{17},X_{21},Y_{21},X_{24},X_{47},X_{51},X_{54},Y_{54},X_{62},X_{65}\})=3\,, \\[.1cm] 
\Delta(\{X_{13},X_{32},Y_{32},X_{35},X_{43},Y_{43}\})=4\,, \\[.1cm] 
\Delta(\{X_{16},X_{75}\})=5\,.
\end{array}
\end{equation}
Note that $\Delta(W)=11$, which is equal to the number of perfect matchings at the origin. Following \cite{2006math......3558B,Eager:2010yu}, we compute the HS using these scalings and find
\begin{equation}
HS=\frac{1+3t^{11}+9t^{22}+3t^{33}+t^{44}}{(1-t^{11})^3(1+t^{11})^2}\,.
\end{equation}
We have explicitly verified that this reproduces several of the first terms in the expansion of the Ehrhart series for the toric diagram calculated with the red origin, providing further confirmation of our prescription.

\subsubsection*{Green Origin}

Choosing the green dot as the origin amounts, $\mathcal{P}_{\mathcal{O}^{\rm green}}=\{p_{12},\ldots,p_{18}\}$, and the reduced $P$-matrix becomes
\begin{equation}
\resizebox{1\textwidth}{!}{$
P^T|_{\mathcal{P}_{\mathcal{O}^{\rm green}}}=
\left(
\begin{array}{c|ccccccccccccccccccc}
 & X_{13} & X_{16} & X_{17} & X_{21} & Y_{21}  & X_{24} & X_{32} & Y_{32} & X_{35} & X_{43} & Y_{43} & X_{47} & X_{51} & X_{54} & Y_{54} & X_{62} & X_{65} & X_{75} & X_{76} \\ \hline
p_{12}  & 1 & 1 & 1 & 0 & 0 & 1 & 0 & 0 & 1 & 0 & 0 & 1 & 0 & 0 & 0 & 0 & 0 & 0 & 0 \\
p_{13}  & 0 & 0 & 0 & 1 & 1 & 1 & 0 & 0 & 1 & 0 & 0 & 1 & 1 & 0 & 0 & 0 & 0 & 0 & 0 \\
p_{14}  & 0 & 0 & 0 & 0 & 0 & 0 & 1 & 1 & 1 & 0 & 0 & 1 & 1 & 0 & 0 & 1 & 0 & 0 & 0 \\
p_{15} & 1 & 0 & 0 & 0 & 0 & 0 & 0 & 0 & 0 & 1 & 1 & 1 & 1 & 0 & 0 & 1 & 0 & 0 & 0 \\
p_{16} & 1 & 0 & 0 & 0 & 0 & 1 & 0 & 0 & 0 & 0 & 0 & 0 & 1 & 1 & 1 & 1 & 0 & 0 & 0 \\
p_{17} & 1 & 1 & 0 & 0 & 0 & 1 & 0 & 0 & 1 & 0 & 0 & 0 & 0 & 0 & 0 & 0 & 0 & 1 & 1 \\
p_{18} & 1 & 0 & 0 & 0 & 0 & 1 & 0 & 0 & 1 & 0 & 0 & 0 & 0 & 0 & 0 & 1 & 1 & 1 & 0 
\end{array}
\right)
$}
\end{equation}
Following our algorithm,
\begin{equation}
\begin{array}{l}
\Delta(\{X_{17},X_{21},Y_{21},X_{32},Y_{32},X_{43},Y_{43},X_{54},Y_{54},X_{65},X_{76}\})=1\, \\[.1cm]
\Delta(\{X_{16},X_{75}\})=2\,, \\[.1cm] 
\Delta(\{X_{47},X_{51},X_{62}\})=4\,, \\[.1cm] 
\Delta(\{X_{13},X_{24},X_{35}\})=5\,.
\end{array}
\end{equation}
As usual, $\Delta(W)=7$, which is the number of perfect matchings at the origin. Using the methods in \cite{2006math......3558B,Eager:2010yu}, we compute the HS with these scalings and find
\begin{equation}
HS=\frac{1+6t^7+t^{14}}{(1-t^7)^3}\,.
\end{equation}
We computed the Ehrhart series with the green origin and found precise agreement, as expected.

\acknowledgments

We would like to thank the Simons Center for Geometry and Physics for its kind hospitality during the completion of this work. S.F. is supported by the U.S. National Science Foundation grant PHY-2412479. D.R.-G is supported in part by the Spanish national grant MCIU-22-PID2021- 123021NB-I00.

\bibliographystyle{JHEP}
\bibliography{mybib}

\end{document}